# Immune System Approaches to Intrusion Detection - A Review

Uwe Aickelin, Julie Greensmith, Jamie Twycross

School of Computer Science, University of Nottingham, UK
uxa,jqg,jpt@cs.nott.ac.uk

**Abstract.** The use of artificial immune systems in intrusion detection is an appealing concept for two reasons. Firstly, the human immune system provides the human body with a high level of protection from invading pathogens, in a robust, self-organised and distributed manner. Secondly, current techniques used in computer security are not able to cope with the dynamic and increasingly complex nature of computer systems and their security. It is hoped that biologically inspired approaches in this area, including the use of immune-based systems will be able to meet this challenge. Here we collate the algorithms used, the development of the systems and the outcome of their implementation. It provides an introduction and review of the key developments within this field, in addition to making suggestions for future research.

## 1 Introduction

The central challenge with computer security is determining the difference between normal and potentially harmful activity. For half a century, developers have protected their systems using rules that identify and block specific events. However, the nature of current and future threats in conjunction with ever larger IT systems urgently requires the development of automated and adaptive defensive tools. A promising solution is emerging in the form of biologically inspired computing, and in particular Artificial Immune Systems (AISs): The Human Immune System (HIS) can detect and defend against harmful and previously unseen invaders, so can we not build a similar system for our computers? Presumably, those systems would then have the same beneficial properties as the HIS such as error tolerance, adaptation and self-monitoring [13].

Alongside other techniques for preventing intrusions such as encryption and firewalls, intrusion detection systems (IDSs) are another significant method used to safeguard computer systems. The main goal of IDSs is to detect unauthorised use, misuse and abuse of computer systems by both system insiders and external intruders [23].

In the following sections, we briefly introduce the areas of IDSs and AISs through the examination of core components and basic definition. The research, development and implementation of immune-inspired IDSs is catalogued, and is presented in terms of the evolving methodology, algorithmic exploration and system implementation details. An overview of this research area is provided, in conjunction with indications for future areas of study.

## 2 Background

This section gives a brief introduction to two distinct fields of study - Intrusion Detection Systems (IDSs) and Artificial Immune Systems (AISs), setting the background to and defining the terminology used in the sections that follow. For a detailed discussion readers should consult [29], [23], [30], [13] and [10].

### 2.1 Intrusion detection systems

IDSs are software systems designed to identify and prevent the misuse of computer networks and systems. There are a number of different ways to classify IDSs. Here we focus on two ways: the analysis approach and the placement of the IDS, although there has been recent work [11] on alternative taxonomies. Regarding the former, there are two classes: misuse detection and anomaly detection [29]. The misuse detection approach examines network and system activity for known misuses, usually through some form of pattern-matching algorithm. In contrast, an anomaly detection approach bases its decisions on a profile of normal network or system behaviour, often constructed using statistical or machine learning techniques. Each of these approaches offers its own strengths and weaknesses. Misuse-based systems generally have very low false positive rates but are unable to identify novel or obfuscated attacks, leading to high false negative rates. Anomaly-based systems, on the other hand, are able to detect novel attacks but currently produce a large number of false positives. This stems from the inability of current anomaly-based techniques to cope adequately with the fact that in the real world normal, legitimate computer network and system usage changes over time, meaning that any profile of normal behaviour also needs to be dynamic [30].

A second distinction can be made in terms of the placement of the IDS. In this respect IDSs are usually divided into host-based and network-based systems [29]. Host-based systems are present on each host that requires monitoring, and collect data concerning the operation of this host, usually log files, network traffic to and from the host, or information on processes running on the host. Contrarily, network-based IDSs monitor the network traffic on the network containing the hosts to be protected, and are usually run on a separate machine termed a sensor. Once again, both systems offer the advantages and disadvantages. Host-based systems are able to determine if an attempted attack was indeed successful, and can detect local attacks, privilege escalation attacks and attacks which are encrypted. However, such systems can be difficult to deploy and manage, especially when the number of hosts needing protection is large. Furthermore, these systems are unable to detect attacks against multiple targets of the network. Network-based systems are able to monitor a large number of hosts with relatively little deployment costs, and are able to identify attacks to and from multiple hosts. However, they are unable to detect whether an attempted attack was indeed successful, and are unable to deal with local or encrypted attacks. Hybrid systems, which incorporate host- and network-based elements can offer the best protective capabilities, and systems to protect against attacks from multiple sources are also under development [30].

## 2.2 Artificial immune systems

The Human Immune System (HIS) protects the body against damage from an extremely large number of harmful bacteria and viruses, termed pathogens. It does this largely without prior knowledge of the structure of these pathogens. This property, along with the distributed, self-organised and lightweight nature of the mechanisms by which it achieves this protection [23], has in recent years made it the focus of increased interest within the computer science and intrusion detection communities. Seen from such a perspective, the HIS can be viewed as a form of anomaly detector with very low false positive *and* false negative rates.

An increasing amount of work is being carried out attempting to understand and extract the key mechanisms through which the HIS is able to achieve its detection and protection capabilities. A number of artificial immune systems (AISs) have been built for a wide range of applications including document classification, fraud detection, and network- and host-based intrusion detection [10]. These AISs have met with some success and in many cases have rivalled or bettered existing statistical and machine learning techniques. AISs can be broadly divided into two categories based on the mechanism they implement: network-based models and negative selection models, although this distinction is somewhat artificial as many hybrid models also exist. The first of these categories refers to systems which are largely based on Jerne's idiotypic network theory [19] which recognises that interactions occur between antibodies and antibodies as well as between antibodies and antigens. Negative selection models use negative selection as the method of generating a population of detectors. This latter approach has been by far the most popular when building IDSs, as can be seen from the work described in the next section.

## 3 Immune system approaches

In this section, we offer an in-depth review of work relating to the application of AISs to the problem of intrusion detection. Initially, we begin by looking at work comparing broad methodological issues, then move on to describe work which compares the efficacy and advantages and disadvantages of individual AIS algorithms within the context of intrusion detection. Finally, we review complete implementations of immune-based IDSs, firstly from the Adaptive Computation Group at the University of New Mexico and then from other researchers.

### 3.1 Methodological issues

Dasgupta and Attoch-Okine [7] compare idiotypic- and negative selection-based approaches to AIS design. Jerne's idiotypic network model [19] is based on idiotypic effects in which antibodies react to each other as well as antigen. In contrast, in the self-nonself model of Forrest et al [34] probabilistic individual antibodies do not interact. The authors consider several applications of AISs, including anomaly detection, fault diagnosis, pattern recognition and computer security. Specifically relating to computer security, they discuss virus detection and process anomaly detection, describing several different approaches. In UNIX processes, changes in behaviour can be detected through short range correlation of process system calls, especially for root processes. Viruses can be detected through detecting changes to files, or through the use of decoys or honeypots, which use a signature-based approach and monitor decoy programs, observe how they were changed, and build signatures from this for main system.

Aickelin et al [1] discuss the application of danger theory to intrusion detection and the possibility of combining research from wet and computer labs in a theoretical paper. They aim to build a computational model of danger theory which they consider important in order to define, explore, and find danger signals. From such models they hope to build novel algorithms and use them to build an intrusion detection system with a low false positive rate. The correlation of signals to alerts, and also of alerts to scenarios, is considered particularly important. Their ideas build on previous work in immunology by Matzinger [28] and work on attack correlation by Kim and Bentley [21]. Their proposed system collects signals from hosts, the network and elsewhere, and correlates these signals with alerts. Alerts are classified as good or bad in parallel to biological cell death by apoptosis and necrosis. Apoptosis is the process by which cells die as a natural course of events, as opposed to necrosis, where cells die pathologically. It is hoped that alerts can also be correlated to attack scenarios. Where these signals originate from is not yet clear, and they will probably be from a mixture of host and network sources. Examples could include traffic normalisers, i.e. a device which sits in the traffic stream and corrects potential ambiguities in this stream, packet sniffers, i.e. a program for collecting live network data, and IDSs such as Snort [32] and Firestorm [26]. The danger algorithm is, however, yet to be specified, as is the correlation algorithm. Whether the system will actively respond to attacks is also not yet clear. Aickelin et al conclude that if this approach works, it should overcome the scaling problems of negative selection, but that a large amount of research still remains to be done. In the future, they intend to implement such a system.

Begnum and Burgess [4] build on previous work by Burgess [6] and combine two anomaly detection models, pH [35] and cfengine [6]. They are motivated by the need to provide a better, automated response mechanism for the pH system, and better detection capabilities for cfengine, as well as the need to collect more detailed data for further research. By the combination of signals from these two systems they hope to provide a more robust, accurate and scalable anomaly detection system. Their approach is to combine the two systems so that pH is able to adjust its monitoring level based on inputs from cfengine, and cfengine is able to adjust its behaviour in response to signals from pH. They discuss the possibility of using the pH/cfengine combination to provide an automated response mechanism which is able to kill misbehaving processes. This work represents an exploration of how to combine the two systems and does not detail any results of experiments, which the authors intend to carry out in the future.

### 3.2 Algorithmic explorations

In [21], Kim and Bentley observe that the HIS is more complex than just negative selection and evaluate this with respect to AISs, investigating performance and scaling related to network intrusion detection. The work builds on the LISYS system [18] and proposed work by Kim and Bentley which incorporates a phenotype into the generation and matching process [22]. They do not describe the overall architecture of the system though similar details are presented in their 1999 paper [24]. TCP packet headers are used based on communications between a LAN (local area network) and the external network and internal LAN communications. TCP (Transmission Control Protocol) is a commonly used network communication protocol. These are derived from a given dataset, and profiles from this data are extrapolated as test and training data. Thirteen self profiles are constructed based on this data and detectors are generated using negative selection against these profiles. The encoding of the detectors contains a number of alleles represented by an arbitrary numerical value. The different alleles on a chromosome are related to different properties of the packet. This range of values is then subject to a clustering algorithm. The similarity between the self strings and incoming strings in the case of the test data is measured using an r-contiguous bit scheme, where the value of $r$ is chosen after estimating the expected number of detectors, detector generation trials and expected false negative rate. A matching activation threshold is derived from the number of detectors generated.

The authors compare their results with the negative selection based system described by Hofmeyr [18]. To generate the self profiles they use the Information Exploration Shootout dataset [15]. This data contains five specified attacks. The profile generator extracts the following information from the dataset: connection identifier, known port vulnerabilities, 3-way handshake details and traffic intensity. The feasibility in terms of time and resources of the negative selection algorithm is then assessed by calculating the time taken to produce the detector set. This is coupled with the number of detectors needed for feature space coverage, and from this the time taken for the generation of a comprehensive detector set is calculated. Additionally, the anomaly detection rate is recorded and analysed. The maximum number of detectors was varied for each attack included in the test profile. Non-self detection rates for the various attacks were recorded as less than 16% so the detector coverage in this case was not sufficient. It was estimated that for an 80% detection rate it would take 1,429 years to produce a detector set large enough to achieve this kind of accuracy, using just 20 minutes worth of data, and $6 \times 10^8$ detectors would be needed. From these results, they conclude that negative selection produces poor performance due to scaling issues on real-world problems. In their opinion other immune-based algorithms, such as clonal selection, need to be used and a better matching function derived. In the future they intend to evaluate both static and dynamic clonal selection algorithms [23].

Dasgupta and Gonzalez [9] are interested in building a scalable IDS and as a step towards this goal investigate and compare the performance of negative and positive selection algorithms. Positive selection is not found in the selection of T-cells in the natural immune systems, whereas negative selection is. They work from a time-series perspective in terms of scalability and changing self, building on work by Forrest [34] and previous work of their own [8]. Their implementation of the positive selection algorithm generates self using training data and time windows. They use a k-dimensional tree, giving a quick nearest neighbour search. At first, they use only one parameter at once, either bytes per second, packets per second or ICMP packets per second. This is then followed by a combination of all three parameters. An alert is generated when values go beyond a threshold. Their negative selection implementation uses real-valued detectors, with self defined as in their positive selection algorithm. A multi-objective genetic algorithm is used to evolve rules to cover non-self, with fitness correlated to the number of self samples covered, area and overlap with other rules. This allows for niching in the multi-objective problem. They define a variability parameter, *v*, as the distance from self still considered normal. This results in one rule for time windows equal to 1 and 25 rules for time window equal to 3. These rules are then used to build detectors. The system has two parameters to set manually: the size of the time window and the threshold.

To test the system they use a small subset of the 1999 Lincoln Labs outside tcpdump datasets [25]: week 1 for training, and week 2 for testing. In their results, they seem to concentrate on only five attacks from that week and see how many of these they can find. Using a combination of all three parameters, all five attacks were detected. A single parameter yielded detection in 3 out of 5 cases. Positive selection needs to store all self samples in memory and is not scalable, but has very high detection rates compared to negative selection, which has a rate of 60% and 80% for window sizes of 1 and 3 respectively, using $\frac{1}{100}$ of the memory of positive selection. Overall, the best detection rates they found were 95% and 85% for positive and negative selection respectively. They concluded that it is possible to use negative selection for IDSs, and that in their time series analysis, the choice of time window was imperative. In the future they intend to use more data to comprehensively test their system.

### 3.3 System implementations — developments by the Adaptive Computation Group, University of New Mexico

**Early work - analysis** The Adaptive Computation Group at the University of New Mexico, headed by Stephanie Forrest, has been instrumental in the develop-ment of intrusion detection systems which employ concepts and algorithms from the field of AISs. Early work from this group is described in Forrest et al [34], and aims to build an intrusion detection system based on the notion of self within a computer system. Their work builds on previous work on an anti-virus system using immune principles [12], and an intrusion detection system called IDES [27]. The system is host-based, looking specifically at privileged processes, and runs on a system which is connected to the network. The system collects information in a training period, which is used to define self. This information is in the form of root user sendmail (a popular UNIX mail transport agent) command sequences. A database of normal commands is constructed and further sendmail commands are examined and compared with entries in this database. The authors consider the time complexity for this operation be $O(N)$ where $N$ is the length of the sequence. A command-matching algorithm is implemented and new traffic compared with the defined behaviour in the database. Intrusions are detected when the level of mismatches with entries in the database becomes above a predefined level. Subsequent alerts are generated but no direct system changing response is implemented.

Building on previous work by the group [34], the work by Hofmeyr et al [16] is also motivated by the need to improve anomaly-based intrusion detection systems. Misbehaviour in privileged processes was examined through scrutinising the same superuser protocols, but using a different representation. System call traces are presented in a window of system calls, a value of 6 in this case. This window is compared against a database of normal behaviour, stored as a tree structure, compiled during a training period. If a deviation from normal is seen, then a mismatch is generated, with sequence similarity assessed using a Hamming distance metric. A sufficiently high level of mismatches generates an alert, but does nothing to alter the system. No user definable parameters are necessary, and the mismatch threshold is automatically derived from the training data.

In all cases the intrusions were detected by the system. The vast majority of the presented results are evidence of the database scaling well, finding the optimum sequence length and setting the mismatch threshold parameters. With regard to false positives, a bootstrap method was used as a proof of concept, though no actual results were presented. The authors conclude that false positives are reduced with an increase in the training period. It is claimed that their system is scalable, and generates on average four false positives per day, although they did not directly compare their system with any other. The results are suggestive that this approach could work using data from both real and controlled environments, but found that it was difficult to generate live data in a dynamic environment. They also note that issues of efficiency have been largely ignored, but will have to be addressed if this is to work in the real world. In the future they intend to perform more fine-grained experiments, and implement a response which is not just based on user alerts, and to incorporate more immune principles.

**Later work - synthesis** The incorporation of some of these suggestions was presented by Hofmeyr and Forrest [17]. The goal of this work was to constructing a robust, distributed, error tolerant and self protecting system. Following on from the previous work of Hofmeyr and Forrest [16] and Forrest et al [34], they aimed to implement and test an IDS based on several different components of the HIS. Their system is network-based and examines TCP connections, classifying normal connections as self, and everything else as non-self. Detectors in the form of binary strings are generated using negative selection, and TCP connections are represented in the form of a data-path triplet, and are subsequently matched against sniffed triplets from the network using an r-contiguous bit matching scheme. If a detector matches a number of strings above an activation threshold, an alarm is raised. Detectors that produce many alarms are promoted to memory cells with a lower activation threshold to form a secondary response system. Permutation masks are also implemented to prevent holes in the self definition. Co-stimulation is provided by a user specifying if an alert is genuine, which reinforces true positives. The activation threshold is set according to an adaptive mechanism involving many local activation thresholds, based on match counts of detectors. Their system is distributed across several machines on the network and therefore one central machine does not have to analyse all the traffic on the network [23]. While the focus of the paper is to describe the algorithms and immune concepts, some experiments are briefly described and it was additionally shown that the rate of false positives can be reduced with user aided co-stimulation.

Following criticism by Kim and Bentley in [21] regarding scaling and false positives, Balthrop et al [3] provide an in-depth analysis of the LISYS immune-based IDS, which evolved from some of the research described above. Their work uses a simpler version of LISYS, a system developed by Hofmeyr [18], in addition to work of Kim and Bentley [22]. Balthrop's system monitors network traffic and is deployed on individual hosts. A detector set is distributed to each of the hosts in the network and TCP connections, based on triplets, are monitored using these detectors. Diversity is created through each host independently reacting to self and nonself. The system uses a negative selection algorithm to mature 49-bit binary detectors which are tested against connections collected during a training period. The matured detectors are then deployed on a live network. An anomaly is detected when a detector has matched a number of connections over a threshold parameter, using an r-contiguous matching function. The generality of the detectors is improved through affinity maturation, and once an intrusion is detected, an alert message is generated. Co-stimulation and permutation masks, both present in the original system, are not implemented. In this case, the user is responsible for setting the value of *r* for the matching function.

Initially, detectors are randomly generated and subject to negative selection so that detectors that match good TCP connections are destroyed. An activation threshold parameter is set automatically by the system, depending upon the number of matches that a detector has made. Additionally, this parameter has a temporal element in the form of a *decay rate*, which is thought to reduce false positive rates. The value for *r* can be varied manually, as can the total number of detectors allowed, the length of the tolerisation period, and the decay rate for the activation threshold. They compared their system, in terms of components but not in terms of performance, to one described by Kim and Bentley [21], and to the US government CDIS system [36] which also uses negative selection.

The experiments took on two parts: the first stage involved defining the best parameters, the second running attacks on the system. In the first part, the number of detectors was investigated, specifically the effects of detector saturation. In the second part, several attacks were performed through the use of Nessus [31]. They found that once the number of detectors reached a certain point then saturation occurred. At this saturation point, the greater the value of *r,* the better the detector set coverage. Balthrop et al also found that the longer the tolerisation period, the fewer the false positives, and that increasing the activation threshold reduced the number of false positives. No information was provided as to the statistical significance of the tests. Overall, detection was successful for the attacks in all but one instance, and the tuning of the parameters reduced the false positive rate. Regarding the scaling issue raised by Kim and Bentley, it was noted that the sensitivity of the system has to be investigated before it can be deployed.

The 'light' version of the LISYS system described in Balthrop et al [3] above was used again in [2], though the focus of this research was on improving the representation of the detectors by exploring a richer representation. The dataset and the experimental system used is the same as in [3]. Their experiments investigate the improvement to r-contiguous matching using an r-chunk scheme. In this scheme, only *r* regions of the whole detector are specified, with the remaining becoming wild-cards. This is thought to reduce the amount of holes that can be present in the detector coverage by the elimination of crossover and length-limited holes during the creation and deployment of detectors.

The effect of permutation masks on the system performance was examined, measured in terms of false positives, and was found to increase the generalisation of the detector coverage. This is based on the observation that an anomaly is likely to produce multiple alerts. Additionally, they found that varying *r* had little effect, unlike with full length detectors. As the r-chunks scheme performed remarkably well the authors investigated it further, and subsequently found that the dramatic increase in performance was in part due to the configuration of their test network. Nevertheless, it still outperformed the full-length detector scheme. They also found that the incorporation of r-chunks and permutation masking reduced false positives and increased true positives. The results of this series of experiments was compared with the setup described in Balthrop et al [3] above. Balthrop et al conclude that r-chunks is appealing as a matching scheme, and that the addition of permutation masks is useful in controlling the rate of false positives. In the future, they intend to run their system on a larger dataset with more attacks.

## 3.4 System Implementations — developments from other researchers

The work performed at the University of New Mexico has contributed significantly to the development of AISs for IDSs. However, they are by no means the only researchers to have actually implemented systems in this manner. This section aims to outline system implementations performed by a number of different research groups, with the common goal of implementing various AISs for applications within security.

The AIS described by Kephart [20] is one of the earliest attempts of applying HIS mechanisms to intrusion detection. It focuses on the automatic detection of computer viruses and worms. As interconnectivity of computer systems increases, viruses are able to spread more quickly and traditional signature-based approaches, which involve the manual creation and distribution of signatures, become less effective. Hence they are interested in creating a system which is able to automatically detect and respond to viruses. Their proposed system first detects viruses using either fuzzy matching from a pre-existing signature of viruses, or through the use of integrity monitors which monitor key system binaries and data files for changes. In order to decrease the potential for false positives in the system, if a suspected virus is detected it is enticed by the system to infect a set of decoy programs whose sole function is to become infected. If such a decoy is infected then it is almost certain that the detected program is a virus. In this case, a proprietary algorithm, not described in the paper, is used to automatically extract a signature for the program, and infected binaries are cleaned, once again using a proprietary algorithm not described in the paper. In order to reduce the rapid spread of viruses across networks, systems found to be infected contact neighbouring systems and transfer their signature databases to these systems. No details of testing and performance are given by the author, who claims that some of the mechanisms are already employed in a commercial product, which other are being tested in a laboratory setting.

In [14], Gonzalez and Dasgupta build an anomaly detector that only requires positive samples, not negative ones, and compare this to a self-organising map (SOM) approach. SOMs are a data dimensionality reduction technique using self-organising neural networks. This work explores the issue of scalability, binary versus real value detectors and a fuzzy distinction between self and non-self, building on work in previous papers by the authors [8] and Forrest [34]. Their system uses real-valued negative selection with n-dimensional vectors as detectors. Detectors have a radius *r*, in other words they represent hyper-spheres. A fuzzy Euclidean matching function is used. In training, detectors are generated randomly and then moved away from self and spaced out. Detectors match if the median distance to their k-nearest neighbours is less than *r*. After a certain time detectors die of old age and eventually a good set of detectors should be found. These detectors are then used to generate *abnormal* samples. A multi-layer perceptron classifier trained with back-propagation is then used to learn to distinguish between self and nonself, after which real data comes in and is classified. Any abnormalities are reported by the system to an operator. They concluded that scaling is not a problem in negative selection when real values are used rather than binary and r-continuous matching. They also concluded that negative selection could train a classifier effectively without seeing non-self. In the future, they intend to use immune networks rather than artificial neural networks.

Le Boudec and Sarafijanovic [5], [33] build an immune-based system to detect misbehaving nodes in a mobile ad-hoc network. These are wireless networks in which each end-user system, termed a node, acts a both a client and router. As nodes act as routers, their proper functioning is essential for the transmission of information across the network. The authors consider a node to be functioning correctly if it adheres to the rules laid down by the common protocol used to route information, in their case the Dynamic Source Routing (DSR) protocol. Each node in the network monitors its neighbouring nodes and collects one DSR protocol trace per monitored neighbour. Even in low bit-rate networks, the amount of routing traffic becomes large and potentially prohibitive in relation to the negative selection algorithm the authors employ. This lead them to adopt a strategy in which DSR protocol events are sampled over a fixed, discrete time intervals to create a series of data sets.

The protocol events within each data set are then reduced, through the identification of four sequences of protocol events. This creates a binary antigenic representation in which each of the four genes records the frequency of their four sequences of protocol events within each data set. The mapping from raw data to antigen was chosen by the authors in such a way that genes within each antigen correlated in a certain way for nodes behaving correctly, and in a different manner for misbehaving nodes. A negative selection algorithm is then used with the generated antigens and a set of uniformly randomly-generated antibodies to eliminate any antibodies which match using a exact matching function. In this maturation stage all collected protocol events are assumed to be indicative of routing traffic between well-behaved nodes. Once a mature set of detectors has been generated, these antibodies are used to monitor further traffic from the node and, if they match antigens from the node, classify it as suspicious.

| system | self-non-self | gene libraries | negative selection | clonal selection | immune memory | idiotypic networks | response |
|---|---|---|---|---|---|---|---|
| Kephart [20] | x | | | | | | x |
| Forrest [34] | x | | | | | | |
| Hofmeyr [16] | x | | | | | | |
| Hofmeyr [17] | x | | x | | x | | |
| Balthrop [3], [2] | x | | x | | | | |
| Gonzalez [14] | x | | x | | | | |
| Le Boudec [5], [33] | x | | | | | | |
| Dasgupta [9] | x | | x | x | | | |

**Table** 1. Summary of immune-based algorithms used by the systems reviewed

# 4 Discussion and Conclusions

The information presented in sections 3.1 to 3.4 has provided detailed overviews of systems which have been implemented, containing one or more immune-inspired algorithms or concepts. In order to clarify the use of various different types of immune algorithm, we shall concentrate here on 'complete' systems, rather than ideas or partial implementations.

Table 1 presents each of the chosen systems and records, in our opinion, which algorithms were used. Within the context of this table we regard an identification in a column as conforming to the following criteria:

- Gene libraries mean the system implemented does not initialise random detector genotypes, but does this through the use of an evolutionary method. –
Negative selection refers to the process of selection of detectors based on elimination if binding to self occurs.
- Clonal selection refers to the B-cell based analogy of increasing detector generality and coverage through the process of hypermutaion.
- Immune memory refers to a secondary response, meaning a similar and more rapid response is elicited should the same attack occur again, irrespective of the time between the attacks.
- Networks correspond with an implementation of the idiotypic network theory, where the different immune components have an effect on each other.
- A response within this context does not simply mean the generation of an alert, but an implemented change in the system as the result of a detection.
- Self-nonself refers to the sense of self, as in the system's recognition of what is normal, or belonging to the system, in order to detect the opposite, that is, nonself.

From Table 1 it is evident that the most popular means of implementing an immune system is through the use of a self-nonself model. This approach is used by all systems under review. Furthermore, negative selection is popular, first used by researchers from New Mexico and then adopted by Dasgupta et al We only found one system each that used the comparatively more advanced features of response, immune memory and clonal selection respectively. No system reviewed used idiotypic networks or gene libraries.

Thus, one can conclude that immunologically inspired IDSs still have much room to grow and many areas to explore, as first observed by Kim and Bentley [21]. Experimental results so far have shown that *relatively simple* AIS based IDSs can work on *relatively simple* problems, i.e. selected test data and small to medium sized testbeds. Will larger scale implementations that borrow more heavily from the HIS, i.e. by incorporating aspects such as idiotypic networks, gene libraries and danger theory, be successful. Such work is currently underway by [1] and others. The proof is yet outstanding, but if it works *in vivo,* we ought to be able to make it work *in silico!*

## Acknowledgements

This project is supported by the EPSRC (GR/S47809/01), Hewlett-Packard Labs, Bristol, and the Firestorm intrusion detection system team.